\pdfoutput=1
\documentclass[letterpaper,twocolumn,10pt]{article}
\usepackage{usenix-2020-09}

\usepackage{algorithm}
\usepackage{algorithmic}
\usepackage{amsmath}
\usepackage{amssymb}
\usepackage{balance}
\usepackage{booktabs}
\usepackage[capitalise,nameinlink,noabbrev]{cleveref}
\usepackage{enumitem}
\usepackage{graphicx}
\usepackage{mathtools}
\usepackage{subcaption}
\usepackage{xspace}
\usepackage[T1]{fontenc}
\usepackage{textcomp}

\usepackage[font=small]{caption}
\usepackage{xcolor}

\usepackage[available,functional,reproduced]{usenixbadges}

\newcommand{\secref}[1]{\S\,\ref{#1}}
\newcommand{\code}[1]{\texttt{#1}}

\DeclareMathOperator*{\argmin}{arg\,min}

\newcommand{\sys}{{Ray Data}\xspace}

\makeatletter
\renewcommand\section{\@startsection{section}{1}{\z@}{-2.5ex plus -1ex minus -.2ex}{1.5ex plus .2ex}{\reset@font\large\bf}}
\renewcommand\subsection{\@startsection{subsection}{2}{\z@}{-2.25ex plus -1ex minus -.2ex}{1ex plus .2ex}{\normalfont\large\bfseries}}
\renewcommand\subsubsection{\@startsection{subsubsection}{3}{\z@}{-2.25ex plus -1ex minus -.2ex}{1ex plus .2ex}{\normalfont\normalsize\bfseries}}
\makeatother
\renewcommand{\paragraph}[1]{\noindent\textbf{#1}}

\newcommand{\op}[1]{\textit{#1}}
\newcommand{\variant}[1]{\textit{#1}}
\newcommand{\inst}[1]{\texttt{#1}}

\setlist[itemize]{leftmargin=1em}

\begin{document}
\date{}

\pagestyle{empty}

\title{\Large \bf Ray Data: A dynamic and scalable system for heterogeneous data processing}
\author{
    Frank Sifei Luan$^{1\,*}$ \enskip
    Ron Yifeng Wang$^{1\,*}$ \enskip
    Yile Gu$^{2}$ \enskip
    Ziming Mao$^{1}$ \enskip
    Charlotte Lin$^{1}$ \enskip
    Amog Kamsetty$^{1}$ \enskip
\\
    Hao Chen$^{3}$ \enskip
    Cheng Su$^{3}$ \enskip
    Balaji Veeramani$^{3}$ \enskip
    Scott Lee$^{3}$ \enskip
    SangBin Cho$^{3}$ \enskip
    Clark Zinzow$^{4}$ \enskip
\\
    Eric Liang$^{5}$ \enskip
    Ion Stoica$^{1}$ \enskip
    Stephanie Wang$^{2}$ \enskip
\\
\\
$^{1}$UC Berkeley \enskip
$^{2}$University of Washington \enskip
$^{3}$Anyscale \enskip
$^{4}$Together AI \enskip
$^{5}$Databricks \enskip
}

\maketitle

\begingroup
\renewcommand\thefootnote{\fnsymbol{footnote}}
\footnotetext[1]{Equal contribution.}
\endgroup

\begin{abstract}
While ML model training and inference are both GPU-intensive, CPU-based data processing is often the bottleneck.
Existing distributed data processing systems assume homogeneous resource requirements.
They excel at CPU-based computation but either under-utilize heterogeneous resources or impose high overheads on failure and reconfiguration.

We introduce \sys{}, a scalable system for heterogeneous workloads spanning batch inference and ML training.
The key idea is to combine \emph{dynamic task creation} for memory efficiency with \emph{dynamic task placement} for operator elasticity.
Making both decisions dynamically allows the system to adapt to heterogeneous pipelines where the memory and compute requirements per operator are unknown and variable over time.
\sys{} addresses two challenges: (1) leveraging dynamic task creation for memory-efficient pipelining, and (2) designing a task placement policy that jointly allocates heterogeneous compute and memory resources to maximize throughput while limiting memory usage.
\sys{} improves throughput on heterogeneous batch inference pipelines by 2.6--12$\times$ compared to traditional data processing systems.
By leveraging heterogeneous clusters, \sys{} improves end-to-end training time for multimodal models such as Stable Diffusion by 31\% compared to popular ML data loaders.
\end{abstract}

\section{Introduction}

Data processing is critical to machine learning (ML) applications.
While ML workloads are known to be GPU-intensive, they also require significant I/O and CPU to load, preprocess, and move data to the GPU.
Indeed, \emph{CPU-based preprocessing is often the bottleneck} in both training~\cite{tfdata} and batch inference~\cite{willump}.
Meanwhile, as ML models evolve, the data processing functionality required has also become more diverse, spanning many data modalities such as text, image, and video.
These new workloads introduce complex transformations and resource requirements~\cite{resnet50,videomae,openai2024gpt4}.
As ML models have grown larger, there is also a critical need to scale the data processing alongside the ML workload.

\begin{figure}[t]
    \begin{subfigure}[b]{0.5\textwidth}
        \centering
        \includegraphics[width=0.5\textwidth,trim=5.9cm 6cm 4cm 5.9cm, clip,page=1]{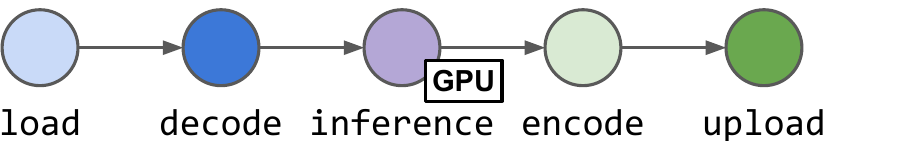}
        \subcaption{Batch inference.}
        \label{fig:model:dag}
    \end{subfigure}
    \begin{subfigure}[b]{0.5\textwidth}
        \centering
        \includegraphics[width=0.6\textwidth,trim=3cm 2cm 6cm 5cm, clip]{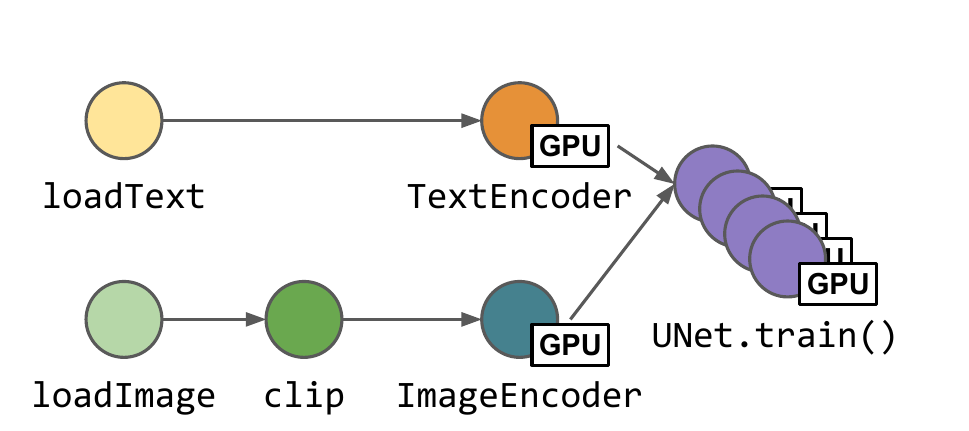}
        \subcaption{Multimodal training with Stable Diffusion~\cite{stablediffusion}.}
    \label{fig:stablediffusion}
    \end{subfigure}
    \caption{
    Logical dataflow graphs representing heterogeneous ML applications. Nodes are operators. \textbf{(a)} A pipeline for video or image generation.
    \textbf{(b)} A multimodal training pipeline. The \code{UNet} model is replicated for distributed data-parallel training.
    }
    \label{fig:applications}
\end{figure}

At first glance, such applications seem simple to scale: both distributed training and batch inference can be expressed as dataflow graphs~(\Cref{fig:applications}) where each node is an \emph{operator} that executes a transform over its incoming data.
The data transforms used are often embarrassingly parallel, e.g., randomly cropping one image for training.
There have been numerous frameworks built to address this exact problem of scaling arbitrary distributed dataflows~\cite{mapreduce,hadoop,spark,isard2007dryad,naiad,flink,kafka,millwheel}.
Unfortunately, none of these are able to fully address the problem of efficient scaling of \emph{heterogeneous} dataflows that use a mix of resources, e.g., GPUs and CPUs.

Resource heterogeneity requires pipelining execution across different operators to maximize overall utilization.
For example, maximizing GPU utilization requires pipelining GPU compute with any CPU and I/O operations needed for data preprocessing, postprocessing, and loading.
This introduces two requirements.

First, \emph{the system must manage memory for intermediate data between operators that require different resources}.
This is challenging because the system must maximize overall throughput while ensuring sufficient overall memory.
If too little intermediate data is buffered, the GPU will idle, wasting valuable resources.
On the other hand, if too much data is buffered, the system may run out of memory or incur expensive overheads from swapping intermediate data to disk.
This is especially important in multimodal applications, which often produce large intermediate outputs, e.g., one video file produces a series of decoded frames.

Second, \emph{achieving efficiency in a heterogeneous setting requires elasticity to keep the overall pipeline balanced}, as the optimal resource allocation per data record and operator may not be known until run time.
For example, in video generation, longer videos will take longer to process, so simple schemes like round-robin assignment of videos to resources can result in poor utilization.
Ideally, the system should also be able to dynamically reallocate each operator's resources based on actual load.
Similarly, in the distributed setting, the system should be able to add and remove physical resources to the cluster with minimal downtime, including in unexpected cases such as node failures.

We present \sys, a scalable system for heterogeneous workloads such as batch inference and ML training.
The key idea is to combine \emph{dynamic task creation} for memory efficiency with \emph{dynamic task placement} for operator elasticity.
We define a task as the execution of an operator over a record batch.
Task creation is the decision of how many records to batch together for operator execution.
Larger tasks amortize per-task system overheads but increase memory footprint.
Task placement is the decision of which worker should execute a given task.
Making both decisions dynamic allows the system to adapt to heterogeneous pipelines where the memory and compute requirements per operator are unknown and variable.

Concretely, \sys{} executes application pipelines at the granularity of tasks.
To limit task memory usage, \sys{} dynamically sizes each task by choosing how many input \emph{partitions}, i.e., record batches, the task should consume.
\sys{} tasks can also produce a dynamic number of output partitions. These can be concurrently consumed by downstream tasks to free intermediate memory as soon as possible.

\sys{} uses a centralized scheduler to dynamically place tasks.
Tasks can be assigned to any worker in the cluster that meets the operator's resource requirements.
This enables instantaneous resource reallocation across operators at partition boundaries.
During execution, \sys{} logs only the task metadata, enabling fast failure recovery via deterministic re-execution of lost tasks, with low run-time overhead.
To maximize \sys{}'s scheduling throughput, we use asynchronous scheduling and a decoupled dataplane.

\sys{} is built on Ray~\cite{ray}, which provides flexible distributed task execution and a distributed dataplane.
However, Ray requires significant extensions, as its scheduler has no visibility into the logical dataflow~(\Cref{fig:applications}).
Thus, we address two primary challenges: (1) extending Ray with dynamic task creation and leveraging it for memory-efficient pipelining, and (2) designing a task placement policy that jointly allocates heterogeneous compute and memory resources to maximize throughput while limiting memory usage.

We evaluate \sys on workloads spanning diverse resource requirements (CPUs, GPUs), storage (local, cloud), and modalities (text, image, video).
\sys outperforms traditional data processing systems such as Spark~\cite{spark} and Flink~\cite{flink} by up to 12$\times$.
\sys matches the throughput of single-node data loaders specialized to ML training such as tf.data~\cite{tfdata} and PyTorch DataLoader~\cite{torchdataloader} while additionally
leveraging disaggregated and heterogeneous preprocessing clusters.
On a Stable Diffusion~\cite{stablediffusion} benchmark, \sys{} improves training time by 31\% by leveraging a pool of 704 CPUs and 72 heterogeneous GPUs.
In summary, we contribute:
\begin{itemize}[itemsep=0pt,parsep=1pt,topsep=2pt]
    \item An architecture combining dynamic task creation and placement for memory-efficient and elastic execution.
    \item A scheduling policy that dynamically maximizes a heterogeneous cluster's compute utilization while enforcing memory limits.
    \item \sys, an efficient, scalable, and fault-tolerant system for heterogeneous ML applications.
\end{itemize}

\begin{figure*}[t]
    \begin{subfigure}[b]{0.29\textwidth}
        \centering
        \includegraphics[keepaspectratio,width=0.98\textwidth,trim=7.5cm 3.5cm 7.5cm 3.5cm, clip,page=5]{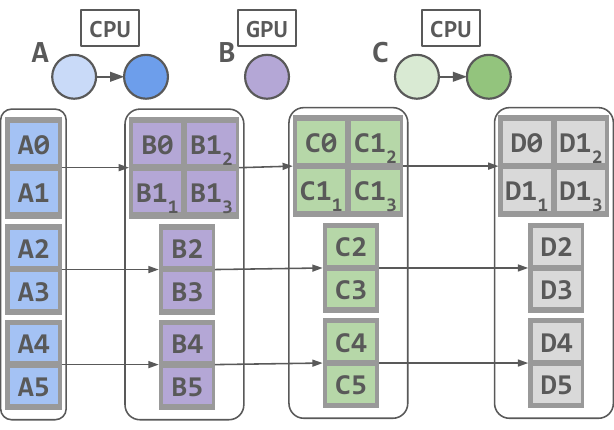}
        \subcaption{Static creation, dynamic placement.}
        \label{fig:system:batch}
    \end{subfigure}
    \begin{subfigure}[b]{0.39\textwidth}
        \centering
        \includegraphics[keepaspectratio,width=0.8\textwidth,trim=5cm 3.7cm 5.5cm 0.5cm, clip,page=6]{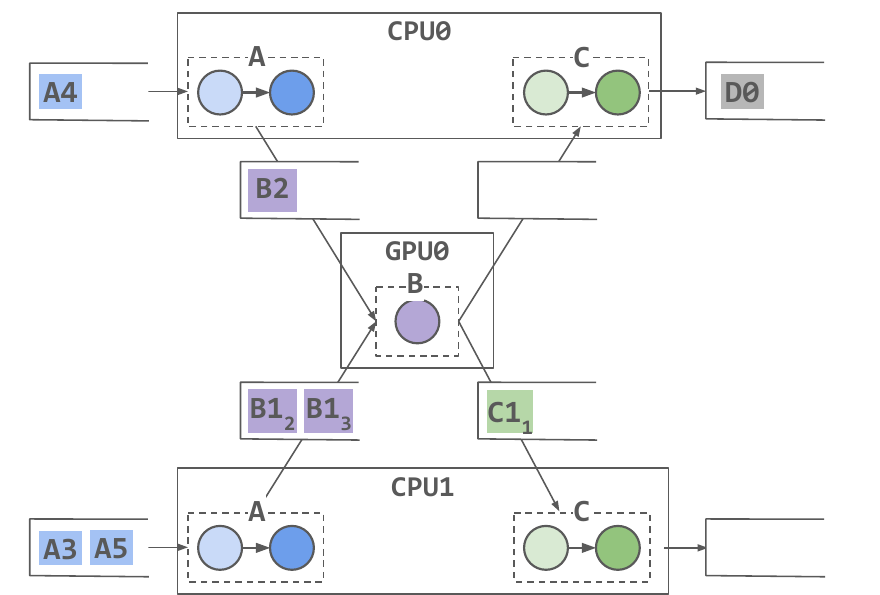}
        \subcaption{Dynamic creation, static placement.}
        \label{fig:system:stream}
    \end{subfigure}
    \begin{subfigure}[b]{0.31\textwidth}
        \centering
        \includegraphics[keepaspectratio,width=0.98\textwidth,trim=6.75cm 3.2cm 6.8cm 3.2cm, clip,page=8]{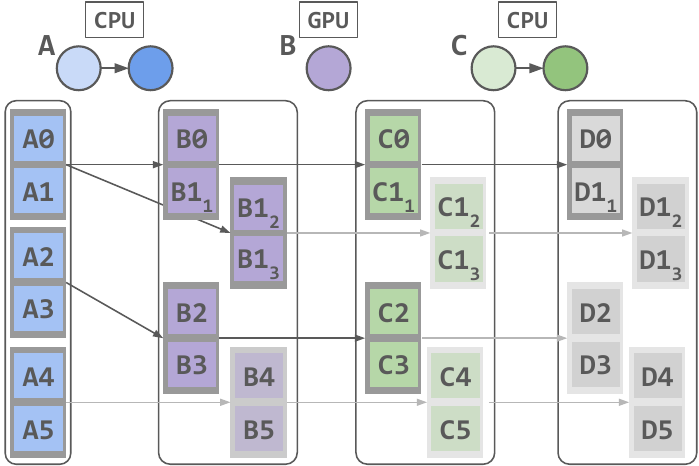}
        \subcaption{Dynamic creation, dynamic placement.}
        \label{fig:system:streamingbatch}
    \end{subfigure}
    \caption{Execution plans for video generation (\Cref{fig:model:dag}). A0, A1, etc. are data records. Consecutive operators that require the same resources are fused, e.g., \texttt{load} and \texttt{decode} are fused into A. A1 is a large video that decodes to 3 video segments, B$1_1$, B$1_2$, B$1_3$.
    \textbf{(a)} Before execution, the system creates tasks by dividing the dataset into evenly sized partitions, e.g., the box with A0 and A1.
    At run time, tasks are assigned resources dynamically.
    \textbf{(b)} Before execution, the system assigns operators to executors and a range partition per operator instance (e.g., even rows for CPU0). At run time, executors dynamically buffer and batch intermediate records.
    \textbf{(c)} Before execution, \sys{} divides only the initial dataset into partitions.
    At run time, each task is dynamically assigned resources and can dynamically produce output partitions. Lighter partitions indicate ones that have not yet been materialized.
    }
    \vspace{-1em}
    \label{fig:system}
\end{figure*}

\begin{figure}[t]
    \centering
    \includegraphics[keepaspectratio,width=0.51\textwidth,trim=2.5cm 0cm 2.5cm 0cm, clip,page=7]{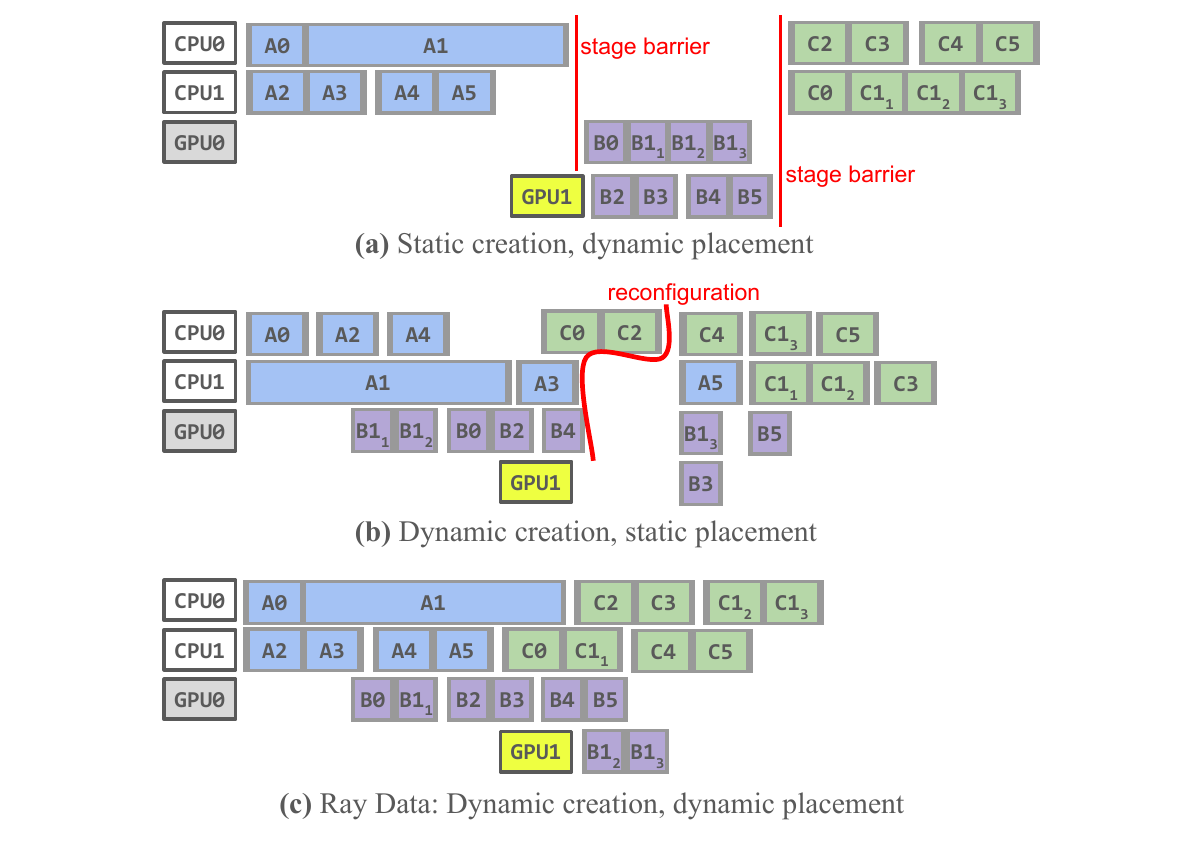}
    {\phantomsubcaption\label{fig:model:a}}%
    {\phantomsubcaption\label{fig:model:b}}%
    {\phantomsubcaption\label{fig:model:c}}
    \vspace{-2em}
    \caption{
    Execution timelines for \Cref{fig:system}. GPU1 is added to the cluster during execution to show how each system handles reconfiguration.
    \textbf{(a)} Tasks are dynamically assigned to resources, so GPU1 can be used immediately. However, tasks execute one stage at a time, and the number of partitions per task must be set before run time.
    \textbf{(b)} Executors locally decide how many records to batch together into a task for better pipelining. Task assignment to executors is static, so cluster reconfiguration can require expensive synchronization.
    \textbf{(c)} \sys{} dynamically creates and assigns tasks, and tasks dynamically produce partitions, producing even-sized partitions.
    }
    \label{fig:model}
\end{figure}

\section{Background}
\label{sec:background}

\subsection{Applications}

We target \emph{throughput} for ML training and inference pipelines that require data pre- and post-processing using CPUs, GPUs, or both.
Similar to current ML data loaders~\cite{tfdata,torchdataloader}, we primarily target map-style per-row transforms.
Operations that require all-to-all shuffle exchanges, such as sort and group-by, are supported through techniques described in \cite{exoshuffle}.

\Cref{fig:model:dag} shows a typical batch inference pipeline, which uses CPUs to load a dataset from cloud storage, GPUs to produce predictions, then CPUs to upload the results.
\Cref{fig:stablediffusion} shows a typical multimodal distributed training pipeline, in this case for Stable Diffusion~\cite{stablediffusion}.
The pipeline uses CPUs to load image-text pairs, GPUs to run inference on frozen \code{Encoder} models, and GPUs to train a \code{UNet} model.
Note that GPU operators can also benefit from heterogeneity and disaggregation: placing the \code{Encoders} on cheaper GPUs can reduce cost by allowing \code{UNet} more GPU memory~(\secref{sec:eval:stablediffusion}).

\subsection{Prior data processing systems}
\label{sec:bg:systems}

We overview previous data processing systems~(\Cref{tab:background}) by analyzing how effectively each system can:
\begin{itemize}[itemsep=0pt,parsep=1pt,topsep=2pt]
    \item Maximize utilization of heterogeneous compute.
    \item Manage memory for intermediate data.
    \item Minimize overheads for cluster failures and re-scaling.
\end{itemize}

We broadly categorize previous systems based on their use of static vs.~dynamic task creation and placement.

\paragraph{Static task creation, dynamic task placement.}
\label{sec:bg:batch}
Some systems combine static task creation and dynamic task placement to achieve fast recovery and operator elasticity.
Examples include Dryad~\cite{isard2007dryad}, MapReduce~\cite{mapreduce}, Apache Hadoop~\cite{hadoop}, Apache Spark~\cite{spark}, and Spark Streaming~\cite{dstream}.

Before execution, the system partitions the dataset and each operator into a predefined number of stateless tasks~(\Cref{fig:system:batch}).
During execution, the system places tasks dynamically on an executor with the matching resources available~(\Cref{fig:model:a}).
Elastic scaling is supported by simply adding or removing executors.
Upon a failure, the system simply re-executes lost tasks; assuming the tasks are deterministic and idempotent, this exactly reconstructs the lost data.

However, such systems typically impose two significant restrictions on execution.
First, they use stage-based execution: each operator must fully execute \emph{before} executing the next~(\Cref{fig:model:a}).
This simplifies scheduling and recovery, as a (re)scheduled task never idles waiting for its inputs, but prevents pipelining and requires materializing all outputs between stages.
While homogeneous operators can be fused into one stage to avoid these overheads, fusion of heterogeneous operators would cause significant under-utilization.

Second, because tasks are static, the system must decide the number of output partitions per task \emph{before} execution.
This can cause out-of-memory (OOM) errors when a task produces more data than expected.
For example, \Cref{fig:system:batch} shows a video generation case where the A1 video is very long.
Under the static partition plan, B$0$, B$1_1$, B$1_2$, B$1_3$ would make a partition twice as large as the others.

Systems such as Spark Streaming~\cite{dstream} and Drizzle~\cite{drizzle} can create ``microbatches'' dynamically, where each microbatch processes a portion of the input, but the tasks within each microbatch remain static.
Also, each microbatch is treated as a separate job that internally still uses staged execution, so pipelining across microbatches is challenging.

\paragraph{Dynamic task creation, static task placement.}
\label{sec:bg:stream}
This combination is often used to minimize task scheduling overheads.
Examples include Naiad~\cite{naiad}, Apache Flink~\cite{flink}, and Spark Continuous Processing~\cite{spark-cont-processing}.

Before execution, the system replicates each logical operator across a set of executors and assigns each replica a range partition of input records~(\Cref{fig:system:stream}).
Different logical operators can share an executor through time slicing or multithreading.
During execution, executors dynamically create ``tasks'' without involving a centralized scheduler:
each operator buffers records at the local executor and decides locally when to send a batch to a downstream operator, determining task boundaries at run time and
pipelining execution across heterogeneous resources~(\Cref{fig:model:b}).
Operators can apply local backpressure for memory efficiency: in \Cref{fig:system:stream}, for the large A1 video, GPU0 applies local backpressure to force CPU1 to buffer B$1_2$ and B$1_3$ until the GPU is available to consume more inputs.

However, static task placement is suboptimal in cases where the load per operator is dynamic.
Cluster reconfiguration is typically costly and complex.
Most such systems require rolling back to a global checkpoint upon any change in task placement, including operator resource reallocation, cluster scale-up or scale-down~(\Cref{fig:model:b}), and unexpected failures.

\paragraph{Dynamic task creation, dynamic task placement.}
Some stream processing systems combine dynamic task creation and placement, similar to \sys{}.
Examples include MillWheel~\cite{millwheel}, Apache Kafka~\cite{kafka}, and Cameo~\cite{cameo}.

However, these systems are not well suited to ML data pipelines.
First, MillWheel and Kafka achieve fast reconfiguration through physical logging of intermediate records, which is expensive and unnecessary for ML pipelines with mostly idempotent computations.
Second, they focus primarily on low-latency processing and do not target high-throughput scenarios with high memory pressure, which is critical for ML data pipelines that produce large intermediate data.
For example, Cameo dynamically assigns tasks to meet user-defined latency deadlines; we found that an emulation of Cameo performs similarly to \sys{} when manually tuned to limit memory consumption, but degrades otherwise.

\paragraph{ML data loaders}. Systems such as tf.data~\cite{tfdata} and PyTorch DataLoader~\cite{torchdataloader} use CPUs to load data onto a local GPU.
They use dynamic task creation and sometimes dynamic placement, but are typically limited to a single node: they launch a fixed pool of worker threads or processes that continuously load, preprocess, and feed data to a local GPU.
Thus, they do not support load-balancing across nodes or the use of heterogeneous nodes.
Also, GPUs are assumed to be sinks, which precludes use cases with CPU-based post-processing~(\Cref{fig:model:dag}) or GPU-based preprocessing~(\Cref{fig:stablediffusion}).

\begin{table*}[t]
    \centering
    \small{

    \begin{tabular}{rccccc}
        \toprule
        System & Dyn task creation & Dyn task placement & Recovery & Rollback & Heterogeneous clusters\\
        \midrule
        Batch systems~\cite{mapreduce,hadoop,spark,isard2007dryad} & $\times$ & $\checkmark$ & Logical log & Task   & $\checkmark$ \\
        Microbatch systems~\cite{dstream,drizzle} & $\times$ & $\checkmark$ & Checkpt+Logical log  & Task   & $\checkmark$ \\
        Naiad~\cite{naiad}, Flink~\cite{flink}       & $\checkmark$ & $\times$ & Checkpt    & Epoch  & $\checkmark$ \\
        MillWheel~\cite{millwheel}, Kafka~\cite{kafka}    & $\checkmark$ & $\checkmark$ & Physical log & Record & $\checkmark$ \\
        PyTorch DL~\cite{torchdataloader}          & $\checkmark$ & $\times$ & None             & Job    & $\times$ \\
        tf.data~\cite{tfdata}             & $\checkmark$ & $\checkmark$ & Checkpt   & Epoch  & $\times$ \\
        \sys{} & $\checkmark$ & $\checkmark$ & Logical log & Task   & $\checkmark$ \\
        \bottomrule
    \end{tabular}
    }
    \caption{
        Systems for heterogeneous data processing. \emph{Dynamic task creation}: the system decides how to batch records between operators. \emph{Dynamic task assignment}: the system dynamically assigns tasks to resources. \emph{Recovery}: the recovery approach. \emph{Rollback}: the minimum amount to re-execute upon failure or cluster reconfiguration. \emph{Heterogeneous clusters}: whether the system can scale using heterogeneous nodes.
    }
    \vspace{-1em}
    \label{tab:background}
\end{table*}

\section{Overview}
\label{sec:overview}

We overview the \sys{} design and the challenges in combining dynamic task creation and placement~(\Cref{fig:system:streamingbatch,fig:model}c).

\paragraph{Challenge 1: Dynamic task creation.}
The throughput and memory footprint of a system are sensitive to the partition size.
Our goal is to achieve memory-efficient pipelining without requiring users to tune a static partition plan.
In particular, we want to allow the user to set only a target partition size in bytes, and the system should automatically partition intermediate data records at run time according to their actual memory usage.
This is challenging because then the number of partitions is not known until run time.
Meanwhile, previous systems using lineage-based recovery, i.e., logical logging, typically require logging the partition plan \emph{before} execution~\cite{mapreduce,spark,dstream}.

\begin{figure}[t]
    \begin{subfigure}[b]{0.238\textwidth}
        \centering
        \includegraphics[width=.74\textwidth,trim=1cm 4.5cm 17cm 3cm, clip,page=3]{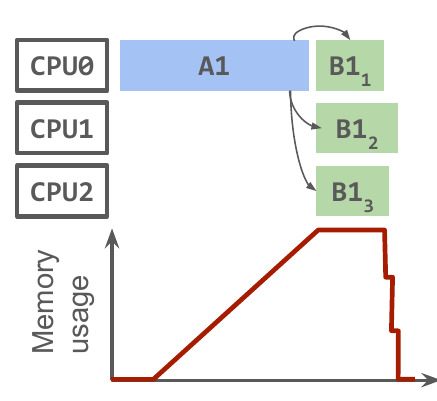}
        \subcaption{Static repartition.}
        \label{fig:repartition:static}
    \end{subfigure}%
    \begin{subfigure}[b]{0.238\textwidth}
        \centering
        \includegraphics[width=.74\textwidth,trim=9cm 4.5cm 9cm 3cm, clip,page=3]{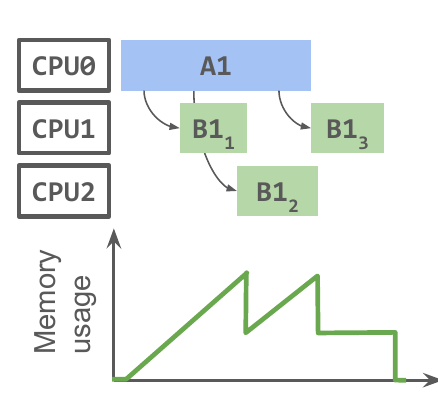}
        \subcaption{Dynamic repartition.}
        \label{fig:repartition:stream}
    \end{subfigure}
    \caption{
    Arrows represent data dependencies.
    \textbf{(a)} Batch systems allow repartitioning, but it must be specified by the user before execution. Execution across operators cannot be pipelined.
    \textbf{(b)} A system with dynamic task creation and placement automatically repartitions outputs to reduce peak memory usage. This also allows the next operator to start in parallel with A1.
    }
    \label{fig:repartition}
\end{figure}

To address this, we propose dynamic partitioning~(\secref{sec:system:repartition}).
The \sys scheduler decides the initial dataset partitioning (dataset A in \Cref{fig:system:streamingbatch}) and maintains a global view of the current status of each partition.
The scheduler dynamically sizes each task by deciding how many partitions it should process. Each task also produces a stream of output partitions.
Each executor locally decides how to partition its outputs based on the target partition size and real-time memory consumption.
For example, in \Cref{fig:repartition:stream}, the A1 task accumulates a buffer of output records.
When the buffer reaches the target partition size, the task flushes the buffer into the B$1_1$ partition.
A1 repeats this until it has outputted all records.
This reduces peak memory usage by allowing B$1_1$ to process and release the intermediate partition concurrently with A1.

To reduce load on the scheduler, we use Ray so that executors only need to return \emph{references} to their output partitions back to the scheduler, not the data~\cite{ownership}.
However, Ray requires tasks to return all outputs at once, and lineage-based recovery requires the number of outputs to be known ahead of time.
We extend Ray's task scheduling and lineage subsystems to allow a task to stream its returned references~(\secref{sec:system:recovery}).
We also design the \sys{} executor to produce dynamic but deterministic output partitions, to ensure recovery correctness.

\begin{figure}[t]
    \begin{subfigure}[b]{0.25\textwidth}
        \centering
        \includegraphics[width=0.9\textwidth,trim=1cm 4.5cm 14cm 3cm, clip,page=4]{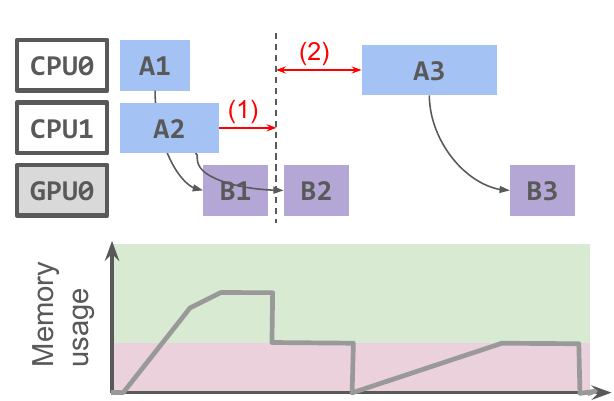}
        \subcaption{Pessimistic scheduling.}
        \label{fig:memory:1}
    \end{subfigure}%
    \begin{subfigure}[b]{0.22\textwidth}
        \centering
        \includegraphics[width=0.82\textwidth,trim=12.25cm 4.5cm 5cm 3cm, clip,page=4]{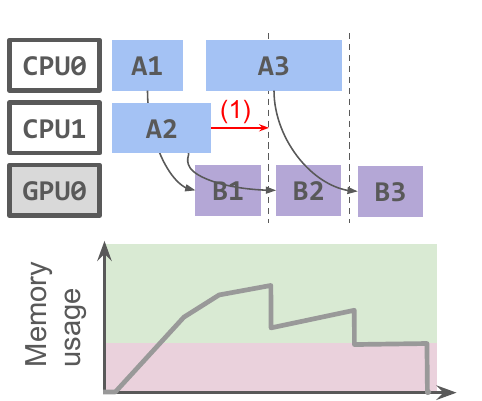}
        \subcaption{Optimistic scheduling.}
        \label{fig:memory:2}
    \end{subfigure}
    \caption{
    Scheduling under memory pressure. Green represents CPU executors' local memory capacity (1 partition per CPU, 2 total). Pink represents the system's shared memory capacity for intermediate data (1 partition total). Arrows represent data dependencies.
    {\color{red} (1)} indicates a period where executors must stall and buffer outputs locally until there is sufficient space in shared memory. \textbf{(a)} further stalls CPU0 during {\color{red} (2)} until there is enough shared memory to schedule \code{A3}, while \textbf{(b)} schedules \code{A3} as soon as possible.
    }
    \label{fig:memory}
\end{figure}

\paragraph{Challenge 2: Dynamic resource assignment.}
The centralized scheduler maintains a global view of the current partition plan~(\Cref{fig:system:streamingbatch}) and the available resources, allowing it to quickly reallocate resources to different operators at partition boundaries and implement cluster-level scheduling policies such as leveraging data locality to reduce data movement for distributed data-parallel training.
However, efficient resource allocation for a heterogeneous pipeline is challenging and requires joint control with memory allocation.

For example, consider \Cref{fig:memory:1}.
Each CPU executor has local memory capacity (green) for 1 partition, and the system has shared memory capacity (pink) for 1 partition.
In phase (1), \code{A2} stalls and buffers its outputs locally until \code{B1} completes.
A conservative scheduler also waits for phase (2) before scheduling \code{A3}, to avoid stalling CPU0.
Meanwhile, \Cref{fig:memory:2} reduces overall run time by scheduling \code{A3} optimistically so that it finishes simultaneously with \code{B2}.
Applying such an optimization successfully would ideally require knowing \code{A3}'s run time and memory footprint.

\Cref{sec:scheduler} introduces our adaptive scheduler, which provides both policies in \Cref{fig:memory}.
We use a conservative backpressure mechanism similar to stream processing systems to implement \Cref{fig:memory:1}.
For \Cref{fig:memory:2}, we use run-time profiling to estimate future memory availability, enabling robust optimistic scheduling.
Because we use a centralized scheduler, both policies are simple to specify.

\begin{table}[t]
    \footnotesize{
    \begin{tabular}{cp{0.34\textwidth}@{}}
    \toprule
        Method & Description \\
        \midrule
        \code{read} & Read items from files. \\
        \midrule
        \code{map} & Transform each item. \\
        \code{map\_batches} & Transform a batch of items. Useful for controlling GPU batch size. \\
        \code{flat\_map} & Transform each item and flatten the results. \\
        \code{filter} & Return items that match a predicate. \\
        \code{limit} & Truncate to the first N items. \\
        \midrule
        \code{write} & Write items to files. \\
        \code{iter} & Return an iterator of items. \\
        \code{iter\_split} & Split into N iterators. \\
        \code{cache} & Materialize all items and cache in memory. \\
    \bottomrule
    \end{tabular}
    }
    \caption{A subset of the \sys{} \code{Dataset} API. The bottom four are consumption APIs that trigger execution, while the others are lazy.\looseness=-1}
    \label{table:api}
    \vspace{0.5em}
\end{table}

\subsection{The Dataset API}

A \code{Dataset} represents an application pipeline.
\code{Datasets} are lazily created, by reading files or applying transforms to another \code{Dataset}~(\Cref{table:api}).
A \code{Dataset} is materialized through a \code{write} to storage, by iterating over the items in memory, or via \code{cache}.

A key part of the API is the ability to express \emph{resource requirements}.
Resource requirements are a map from resource name to float value and may be passed as an option to each transform.
By default, each transform requires 1 CPU.
Resource names can be \code{CPU}, \code{GPU}, or a custom resource label.

The map transforms take a stateless user-defined function (UDF) as an argument.
For operations that require significant initialization time, such as a model loaded into GPU memory, we also support UDFs with read-only state.
Like other lineage-based systems, we assume that UDFs are pure.

\begin{figure}[t]
    \centering
    \includegraphics[width=0.45\textwidth,trim=4cm 3cm 4cm 3cm, clip,page=2]{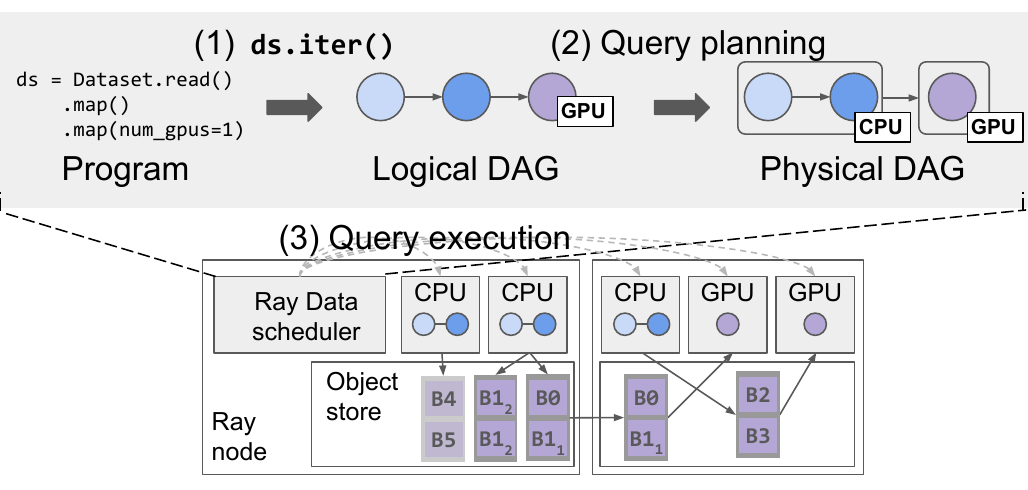}
    \caption{\sys{} architecture overview. \sys{} executes as a Ray library. The \sys{} scheduler maintains the partition metadata~(\Cref{fig:system:streamingbatch}) and dispatches tasks to Ray workers (dashed arrows).}
    \label{fig:overview}
\end{figure}

\subsection{Executing a \sys{} Program}

After creating a \code{Dataset}, the user triggers execution by calling one of the consumption APIs, as seen in (1) in \Cref{fig:overview}.
The \code{Dataset} is represented as a DAG of logical operators~(\Cref{fig:applications}).
The system's query planner then compiles this logical DAG into a DAG of physical operators~((2) in \Cref{fig:overview}).
The query planner applies operator fusion and decides the number of partitions to use for the first operator~(\secref{sec:system:planning}).

\sys{} uses Ray as a task backend, storing intermediate data partitions in Ray's distributed object store.
During execution, the \sys{} scheduler maintains a global view of the current partition metadata~(\Cref{fig:system:streamingbatch}), including a reference to the remote data.
The scheduler then dispatches each physical operator as a series of Ray tasks ((3) in \Cref{fig:overview}). Stateless UDF tasks are scheduled directly by Ray and can run on any Ray worker.
For each stateful UDF, \sys{} creates a pool of Ray ``actors'': stateful workers that acquire resources for their lifetime.
If multiple stateful UDFs require the same resource, multiplexing is possible by sharing an actor pool between UDFs.
Note that while actors hold read-only application state such as a GPU model, they do not store any system state between tasks, so any stateful UDF task can execute on any actor in the pool.
This enables simple load balancing across workers.
Furthermore, reallocating an operator's resources and scaling the cluster up or down simply requires updating the \sys scheduler's map of available resources.

The \sys{} scheduler runs a continuous loop alternating between waiting for a dispatched task to materialize an output partition and release its resources, and choosing a new task to run on any available resources.
In principle, one could instead rely entirely on Ray's scheduler by submitting the entire task graph in \Cref{fig:system:batch} upfront.
However, this would prevent dynamic partitioning, and naive application of Ray's scheduler results in poor performance~(\secref{sec:eval:mb:scalability}), as Ray cannot easily implement the application-aware policies that we present in \Cref{sec:scheduler}.

\section{System Design}
\label{sec:system}

\subsection{Query Planning}
\label{sec:system:planning}

The query planner transforms a logical DAG~(\Cref{fig:applications}) into a physical DAG~(\Cref{fig:system:batch}), then applies optimizations such as operator fusion and the initial partitioning for \code{read}.
We choose enough partitions to use all available execution slots (usually CPUs), but not so many that each partition is tiny, which can increase system overheads.
If the estimated output size is known for a particular file type,  then we also aim to produce partitions that are 1--128\,MB in size~(\secref{sec:eval:mb:partitioning}).
Note that the system is robust to the initial number of partitions because we also dynamically repartition tasks at run time.

\code{Dataset} consumption APIs~(\Cref{table:api}) are also incorporated into the physical DAG.
The \code{Dataset.write} call is implemented with \code{map}.
The \code{iter} API returns a stream of output records.
This is implemented by fetching and buffering materialized output partitions. The \code{iter\_split} call shards the outputs into N streams, each of which can be passed to a different reader process.
This is useful for cases such as distributed data-parallel training where the dataset is sharded among N trainers.
To implement \code{iter\_split}, the query planner uses an additional Ray actor to coordinate dynamic assignment of materialized output partitions to readers.

\subsection{Query Execution}
\label{sec:system:exec}

During execution, the \sys{} scheduler maintains a global view of materialized partitions~(\Cref{fig:system:streamingbatch}), executing tasks, and available resources.
For each partition, the \sys{} scheduler maintains the following metadata: number of rows, size in bytes, and node location (used for data locality).
For each operator, the scheduler maintains a queue of input partitions, stored as Ray references.
It then repeatedly executes this loop:

\begin{itemize}[itemsep=0pt,parsep=2pt,topsep=2pt]
  \item Wait for an executing task to materialize an output partition. Push the partition reference onto the downstream operator's input queue. Tasks may produce multiple output partitions; if this was the last, free the task's resources.
  \item While there are free resources and queued input partitions, select a physical operator to run using the policy described in \Cref{sec:scheduler}. Launch a new task for that operator. Mark the task's required resources as used.
\end{itemize}

To launch a new task, the scheduler passes a closure of the physical operator to execute and references to the task's input partition(s).
Ray ensures that all references are replaced by the physical data on the task executor.
Typically, the scheduler will delete a reference as soon as its downstream task is submitted, signaling to Ray that the data can be garbage-collected after task execution.
To implement \code{Dataset.cache()}, the scheduler simply keeps a reference to all output partitions of the corresponding operator.

\subsubsection{Dynamic repartitioning}
\label{sec:system:repartition}
The query planner's initial estimate of the number of partitions to use may not be optimal.
\sys{} uses dynamic repartitioning to handle such cases.
To support this, we extend Ray with \emph{generator tasks}, enabling remote tasks to produce a dynamic number of outputs, and to pipeline execution with the task's caller (the \sys{} scheduler).
Whenever the task yields a new output partition, it notifies the \sys scheduler via RPC.
Upon receipt, the scheduler reruns the scheduling loop and can launch a downstream task.
Meanwhile, the upstream task continues producing its next output.

\sys{} executors take a target partition size from the scheduler and determine locally how to partition task outputs.
A \sys{} task is an iterator that applies the given transform over its input partition(s).
The executor yields from this iterator until the accumulated buffer exceeds the maximum target partition size (128\,MB by default).
Then, the \sys{} task flushes its buffer as a partition in Ray's object store.
For example, for a video decoding transform, \sys{} could assign one video per task, and each task would yield a stream of video frames.
The \sys{} executor would then dynamically batch the frames into a stream of output partitions.
Once the task finishes, all corresponding system state is removed from the executor.

Operators that produce much less data than they consume can produce too-small partitions.
To handle this, the \sys{} scheduler coalesces partitions by passing multiple partitions from the upstream operator to a single task, up to the maximum target partition size.

For ML workloads, users often want control over GPU batch size.
Thus, we expose a \code{map\_batches} API that accepts a batch transform function and a target batch size.
\sys{} tasks ensure the batch size by iterating over slices of the input partition(s).
The \sys{} scheduler also coalesces too-small partitions to the desired batch size.

\subsubsection{Failure recovery}
\label{sec:system:recovery}

Ray provides automatic recovery for objects (intermediate partitions) as long as (1) the driver is alive, (2) the tasks that created them are deterministic and side-effect-free~\cite{ownership}, and (3) the task arguments and outputs are immutable.
For generator tasks, (3) is no longer true, because we do not know at submission time how many outputs the task will produce.
However, we note that if (2) is true, then dynamic repartitioning can be made deterministic.
In particular, given a target partition size and a pure transform, we ensure that a \sys{} task will produce the same stream of output partitions if executed on the same input partition(s).

To support failure recovery for generator tasks, we modify Ray's recovery subsystem to handle tasks with an unknown number of outputs.
On the first successful execution of a generator task, the task's caller records the number of outputs that the task produces.
If any of the task's outputs are lost, we recover by re-executing the entire task.
If the task produces a different number of outputs, we throw an error.

Similar to other batch and stream processing systems~\cite{flink, paszke2019pytorch, spark, tfdata}, if the centralized scheduler dies, Ray garbage-collects the job and so \sys{} must re-execute from the beginning.
Rollback can be reduced via checkpointing.

\subsection{Scheduler Policy}
\label{sec:scheduler}

The scheduler policy takes as input:

\begin{itemize}[itemsep=0pt,parsep=1pt,topsep=2pt]
    \item The physical DAG of operators~(\Cref{fig:overview}), each annotated with a resource requirement, e.g., \code{\{GPU:1\}}.
    \item Cluster resources: for each node, the number of CPU, GPU, or custom resource slots, and the memory capacity of the shared memory pool that stores partitions.
\end{itemize}

The scheduler policy decides the current mapping between cluster resources and operator tasks/partitions.
Each time a task completes, the policy can allocate the freed resources to a new operator task.
The policy can also choose to allow more than one outstanding task per cluster resource.
This is beneficial for cases such as large language model (LLM) inference, which batches continuously across sequences for better throughput~\cite{yu2022orca}.
With multiple tasks in flight to an LLM replica, batching can be applied to sequences \emph{across} partitions in addition to \emph{within} a partition~(\secref{sec:eval:rag}).

The scheduler enforces the shared memory capacity as a hard limit, using either policy in \Cref{fig:memory}.
We implement the pessimistic policy~(\Cref{fig:memory:1}) by prioritizing operators with shorter output queues and stalling in-progress tasks when the global memory limit has been reached.
This is similar to the backpressure mechanism used in streaming systems.
Next, we describe the optimistic policy that minimizes stalling based on run-time profiling.

\subsubsection{Optimistic policy}
\label{sec:scheduling:impl}

The optimistic policy~(\Cref{fig:memory:2}) aims to minimize stalling and therefore job completion time by estimating when memory will become available in the \emph{future}.
To achieve this, the scheduler estimates and then equalizes the processing rates of each operator (in bytes per second).
Intuitively, if processing rates are not equal, slower operators will accumulate pending inputs, eventually exhausting memory.
The optimistic policy estimates operator processing rates online using run-time statistics: task durations and ratio of input:output size.
The policy controls operator processing rates by deciding how many tasks to run in parallel for that operator.

\begin{algorithm}[t]
\caption{Adaptive Scheduler}
\label{alg:scheduler}
\begin{algorithmic}[1]
\STATE Initialize $budget \gets totalMemoryCapacity$
\WHILE{not all operators are done}
    \STATE Update resource utilization and run-time estimates
    \STATE Update budget \COMMENT{\Cref{alg:budget}}
    \IF{$budget \geq \mathrm{outputPartitionSize}(source)$}
        \STATE Launch task of $source$
        \STATE $budget \gets budget - \mathrm{outputPartitionSize}(source)$
    \ENDIF

    \STATE $Q \gets \emptyset$ \COMMENT{Set of qualified operators}
    \FOR{each operator $op$ in DAG}
        \IF{hasInputData($op$) \textbf{and} \\
            \hspace{1.5em}hasAvailableResources($op$) \textbf{and} \\
            \hspace{1.5em}hasOutputBufferSpace($op$)}
            \STATE $Q \gets Q \cup \{op\}$
        \ENDIF
    \ENDFOR
    \IF{$Q \neq \emptyset$}
        \STATE $selected \gets \argmin_{op \in Q} \text{bufferedOutputsSize}(op)$
        \STATE Launch task of $selected$
    \ENDIF
\ENDWHILE
\end{algorithmic}
\end{algorithm}

\Cref{alg:scheduler} describes the scheduling loop for the optimistic policy.
Like the pessimistic policy, the optimistic policy prioritizes the operator with the least amount of buffered outputs.
However, when memory is constrained, the policy (without lines 4--8) will result in resource under-utilization, as seen in \Cref{fig:memory:1}.
The ideal solution is to keep the pipeline full and start \emph{source} tasks, i.e., tasks for the first operator, as early as possible, as in \Cref{fig:memory:2}.
To achieve this, lines 4--8 add a higher-priority optimistic policy for scheduling source tasks, described next.

The input rate, i.e. the rate at which source tasks are scheduled, must approximate the pipeline's overall throughput.
We use a dynamic \emph{memory budget} algorithm to regulate the rate at which source tasks are launched.
Intuitively, the budget is an optimistic estimate of the memory available for new data partitions to enter the system.
When a source task is launched, we deduct its estimated output size from the budget.
At every second, the budget is replenished using \cref{alg:budget}, which estimates the rate at which data leaves the pipeline.
For DAGs with multiple sources, we launch each source operator at a rate proportionate to its output size.

\begin{algorithm}[t]
\caption{Memory Budget Update (runs every second)}
\label{alg:budget}
\begin{algorithmic}[1]
\STATE $P \gets 0$ \COMMENT{Total processing time per partition}
\STATE $\alpha_0 \gets 1$ \COMMENT{$\alpha_i:=$ Input:Output size ratio for $op_i$}

\FOR{$i \gets 1$ to $numOps$}
    \STATE $E_i \gets \text{availableExecutionSlots}(op_i)$
    \STATE $T_i \gets \text{estimatedTaskDuration}(op_i)$
    \IF{$op_i$ is not $source$}
        \STATE $I_i \gets \text{estimatedInputSize}(op_i)$
        \STATE $O_i \gets \text{estimatedOutputSize}(op_i)$
        \STATE $\alpha_i \gets \alpha_{i-1} \cdot O_i/I_i$
    \ENDIF
    \STATE $P_i \gets (T_i/E_i) \cdot \alpha_{i-1}$
    \STATE $P \gets P + P_i$
\ENDFOR

\STATE $budget \gets budget + \mathrm{outputPartitionSize}(source)/P$
\end{algorithmic}
\end{algorithm}

We will walk through the algorithm using the following example: \op{load} (CPU) $\to$ \op{transform} (CPU) $\to$ \op{inference} (GPU) pipeline, running on a cluster with 8 CPUs and 4 GPUs.
\begin{itemize}[itemsep=0pt,parsep=1pt,topsep=2pt]
    \item Consider the first non-source operator: \op{transform}. Suppose the current number of available execution slots to run the task is $E_1 = 6$ (out of 8 CPU slots).
    Assume the average task duration is $T_1 = 12$ seconds. Then the processing time of this stage is $P_1 = T_1 / E_1 \cdot \alpha_0 = 12 / 6 \times 1 = 2$, where $\alpha_0$, the output multiplier, is initialized to 1.
    In other words, \op{transform} takes 2s to process a source partition on average.
    \item Assume the \op{transform}'s average output is double the size of its input, i.e. $\alpha_1 = 2$.
    Now consider \op{inference}.
    Assume the number of available execution slots is $E_2 = 4$ (GPU), and the average task duration is $T_2 = 2$ seconds. Then $P_2 = T_2 / E_2 \cdot \alpha_1 =  2 / 4 \times 2 = 1$, i.e. the \op{inference} operator takes 1 second per source partition.
    \item Adding them up, $P = 2 + 1 = 3$ seconds per source partition. In other words, every \textasciitilde3s, the budget will be replenished to allow for one more source task to run.
\end{itemize}

If the run-time estimates of each operator's processing rates are perfectly accurate, i.e. if there is no variance in the processing rates, it can be shown that the schedule produced is optimal.
However, when there is variance in processing times or output sizes, the budget algorithm could overestimate the overall processing rate.
Nevertheless, the algorithm is stable because it creates a negative feedback loop.
If it overestimates the pipeline processing rate, more source tasks might launch and temporarily cause backpressure, or objects in the buffer to spill to disk.
However, because these tasks occupy execution slots, they will reduce the parallelism of downstream operators and lower the replenishment rate of the budget, which in turn limits the source task launch rate.

\subsection{Implementation}
\label{sec:impl}

\sys{} is an industrial system and is written in \textasciitilde51k Python LoC, with about 1k, 1k, and 2k LoC for the query planner, scheduler, and map executor logic, respectively (the remainder is user-facing APIs, metrics collection, etc.).
We implement \sys{} on top of Ray because Ray exposes a low-level execution model based on dynamic task execution.
It also provides useful features such as transparent data movement, lineage-based recovery~\cite{ownership}, and disk spilling~\cite{exoshuffle}.
This makes it convenient to build centralized schedulers like \sys{} while leveraging Ray as a decentralized dataplane.
Ray provides cluster resource information, while \sys{} tracks operator buffer usage and maintains per-operator
statistics on task durations and input:output size ratios
for scheduling~(\secref{sec:scheduling:impl}).
Ray's first-class Python support lets ML developers swap in \sys{} for Python-based data loaders with minimal code changes and without moving data across language runtimes, while Ray's C++ dataplane ensures high performance for throughput-oriented workloads.

In principle, it is also possible to build \sys{} by extending a previous data processing system.
For example, Spark could be extended with dynamic task creation, Flink with dynamic resource assignment, or Kafka with logical logging of intermediate data.
In practice, \sys{} is likely easier to port to low-level task or actor systems analogous to Ray~\cite{dask,bernstein2014orleans,srirama2021akka}.

\section{Evaluation}

\begin{figure*}[ht]
  \centering
  \begin{subfigure}[b]{169.64bp}
      \includegraphics[scale=0.681526,trim=0 -4.269 0 -4.151,clip]{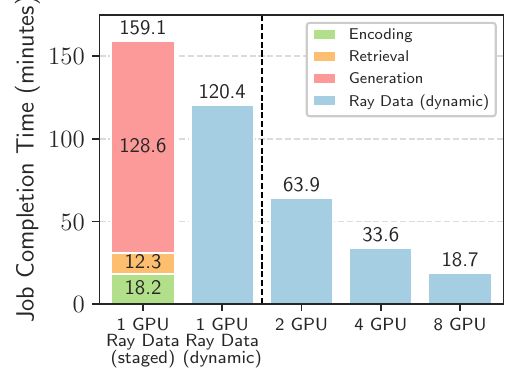}
      \caption{Retrieval-augmented generation.}
      \label{fig:eval:rag}
  \end{subfigure}\hfill
  \begin{subfigure}[b]{166.64bp}
      \includegraphics[scale=0.681526,clip]{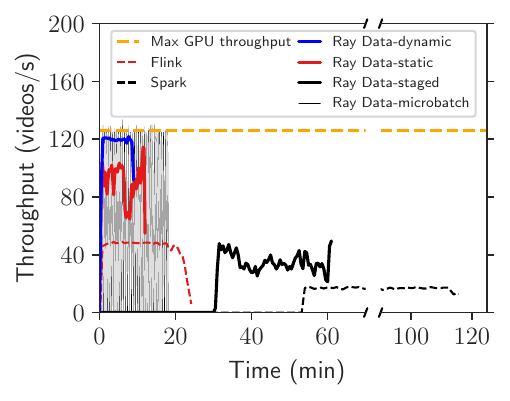}
      \caption{Video classification.}
      \label{fig:eval:video-inference}
  \end{subfigure}\hfill
  \begin{subfigure}[b]{160.26bp}
      \includegraphics[scale=0.681526,trim=0 -1.21 0 0,clip]{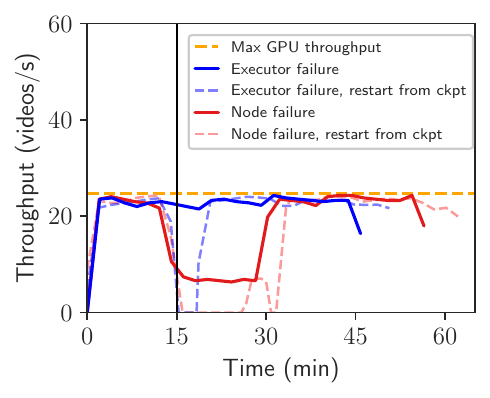}
      \caption{Fault tolerance.}
      \label{fig:video:tput}
  \end{subfigure}
  \caption{
  \textbf{(a)} RAG job completion time. \variant{\sys{}-staged} runs stages sequentially in one process; its \op{Retrieval} run time is the optimal JCT if CPU is the bottleneck.
  \textbf{(b)} Video classification throughput comparison of batch~(Spark and \variant{\sys{}-staged}), streaming~(Flink and \variant{\sys{}-static}), microbatch~(\variant{\sys{}-microbatch}), and dynamic task creation and assignment~(\variant{\sys{}-dynamic}) systems.
  \textbf{(c)} Fault tolerance comparison for video classification on a heterogeneous cluster with 1 GPU node and 1 CPU-only node. Executor (\textcolor{blue}{blue}) or node (\textcolor{red}{red}) failure is injected at $t$=15min. For node failure curves, the node is added back to the cluster at $t$=30min. The dashed line is a variant of \sys{} that emulates checkpoint-and-restore. This strategy leads to job downtime after every cluster change and longer completion time overall.
  Unmodified \sys{} (solid) smoothly handles executor failure and node failure and addition.
  }
  \label{fig:evaluation-results}
  \vspace{-1em}
\end{figure*}

We evaluate heterogeneous workloads in batch (offline) inference and training, including retrieval-augmented LLM generation (RAG) and multimodal models.
We use benchmarks from the MLPerf suite~\cite{mlperf-training,mlperf-inference} and supplement them with additional LLM, image, and video benchmarks.
We study:

\begin{itemize}[itemsep=0pt,parsep=1pt,topsep=2pt]
    \item \secref{sec:eval:inference}: How does \sys{} compare to existing data processing systems when running heterogeneous ML workloads, in terms of throughput and elasticity?
    \item \secref{sec:eval:training}: How can distributed and heterogeneous execution reduce cost compared to single-node training data loaders?
    \item \secref{sec:eval:mb}: How well can all systems adapt to memory pressure from intermediate data in heterogeneous settings?
\end{itemize}

\noindent We compare the following systems:
\begin{itemize}[itemsep=0pt,parsep=1pt,topsep=2pt]
    \item Static creation, dynamic assignment: Apache Spark 3.5.1.
    \item Dynamic creation, static assignment: Apache Flink 1.19.
    \item Dynamic task creation and assignment: \sys{} (implemented over Ray 2.40.0).
    \item Single-node ML data loaders: tf.data~\cite{tfdata} and PyTorch DataLoader (PyTorch DL)~\cite{torchdataloader}. tf.data can reconfigure the number of threads per operator.
\end{itemize}

We also modify \sys{} to emulate the execution models of baseline systems for apples-to-apples comparison. We emulate: (1) staged execution similar to Spark (\variant{\sys{}-staged}), (2) microbatch processing similar to Spark Streaming~\cite{dstream} (\variant{\sys{}-microbatch}), and (3) static task assignment similar to Flink (\variant{\sys{}-static}). \variant{\sys{}-staged} materializes each stage before starting the next. \variant{\sys{}-microbatch} groups inputs into microbatches, but still executes a barrier between heterogeneous operators (e.g., CPU to GPU). \variant{\sys{}-static} sets a static parallelism per operator and replaces the dynamic scheduler~(\secref{sec:scheduler}) with round-robin partition assignment.

For training workloads, we compare only against tf.data and PyTorch DataLoader because these systems were custom-built for data preprocessing for training.
For batch inference, we use Spark, Flink, and \sys{} variants.

\subsection{Inference: \sys{} vs.~batch and stream processing}
\label{sec:eval:inference}

\subsubsection{Retrieval-Augmented Generation (RAG)}
\label{sec:eval:rag}

RAG~\cite{rag} improves LLM inference accuracy by retrieving relevant documents from a knowledge base to augment the original prompt.
Our RAG pipeline consists of three stages:
(1) \op{encode} (CPU): Encode the input prompt using a pre-trained encoder (Contriever~\cite{contriever}) to produce a dense embedding.
(2) \op{retrieve} (CPU): Use a FAISS~\cite{douze2024faiss} vector index to retrieve the top-$k$ most similar documents from the knowledge base (TriviaQA~\cite{joshi-etal-2017-triviaqa}).
(3) \op{generate} (GPU): Serve the Llama-3-8B model~\cite{llama3} with vLLM~\cite{vLLM} to generate the final output based on the input prompt and retrieved documents.
We run the RAG pipeline on 1 node with 8 H200 GPUs and 256 CPUs.

\Cref{fig:eval:rag} shows the job completion time (JCT) for 100K prompts.
We compare our system (\variant{\sys{}-dynamic}) to a single-process baseline \variant{\sys{}-staged} that sequentially executes each stage.
On 1 GPU, \variant{\sys{}-dynamic} achieves 1.32$\times$ speedup compared to the baseline.
This is because \variant{\sys{}-dynamic} runs all stages concurrently and overlaps the CPU- and GPU-based stages, so its overall JCT is similar to that of the \op{generate} stage in \variant{\sys{}-staged}.

We also scale \sys{} to multiple GPUs.
On N GPUs, we use \sys{} to set the parallelism of \op{generate} to N, creating N vLLM replicas.
As shown in \Cref{fig:eval:rag}, compared to 1 GPU, \variant{\sys{}-dynamic} is able to achieve 1.88$\times$ speedup for 2 GPUs, 3.58$\times$ speedup for 4 GPUs, and 6.44$\times$ speedup for 8 GPUs.
Larger overhead is observed when scaling from 4 to 8 GPUs because the CPU-based stages become the bottleneck.

\paragraph{Takeaways:} (1) \sys{} outperforms staged execution by pipelining across heterogeneous resources, and (2) \sys{} scales LLM inference via data parallelism while saturating the resource bottleneck, either GPU or CPU.

\subsubsection{Video Classification}
\label{sec:eval:video}
We run the VideoMAE video classification model~\cite{videomae} on the test split of the Kinetics-700-2020 dataset (64,535 videos, 137.3 GB, Amazon S3), on a 4-node cluster (one \inst{g5.4xlarge} with 16 vCPUs and three \inst{g5.2xlarge} with 8 vCPUs, each with 1 NVIDIA A10G GPU).
The operators are: (1) \code{read}: download and read binary files from S3, (2) \code{preprocess}: decode a video into a series of frames, (3) \code{map\_batches(VideoMAE)}: classify video frames using a GPU-hosted model.
\Cref{fig:eval:video-inference} shows throughput over time.
We include \variant{\sys{}-staged}, \variant{\sys{}-static}, and \variant{\sys{}-microbatch} emulations.

Staged systems~(Spark and \variant{\sys{}-staged}) execute heterogeneous stages synchronously, so they do not produce results until $t$=53min and $t$=31min for Spark and \variant{\sys{}-staged}, respectively.
Also, the JCTs are $t$=116min and $t$=61min respectively because all intermediate stage results are materialized and must be spilled to disk to avoid OOM.

Systems with static task assignment~(Flink and \variant{\sys{}-static}) pipeline across heterogeneous operators so they are able to produce results almost immediately.
However, they are also sensitive to the static assignment of operator ranges to executors.
Both use a round-robin approach to assign data to executors.
Flink's throughput is more stable than \variant{\sys{}-static}'s until $t$=20min, likely due to finer-grained partitions.
However, Flink also has significant overheads from serializing and copying data between Java and Python.
Thus, we also include \variant{\sys{}-static} for fair comparison.
\variant{\sys{}-static}'s throughput is unstable due to lack of dynamic load balancing.

Microbatch systems~(e.g., Spark Streaming~\cite{dstream}, Drizzle~\cite{drizzle}) can decide the size of each microbatch dynamically.
This improves upon staged systems because microbatches can be sized to fit in cluster memory, but each microbatch still internally uses static tasks and staged execution.
We emulate this execution model (\variant{\sys{}-microbatch}) and find that it produces spiky throughput and suffers from barrier synchronization overheads, as the GPU must sit idle while waiting for the CPU stage to finish.
Furthermore, optimizations proposed by systems such as Drizzle~\cite{drizzle} primarily target scheduling latency, which is already in the milliseconds for \variant{\sys{}-microbatch}.
Thus, they have no impact on \variant{\sys{}-microbatch} throughput.

\variant{\sys{}-dynamic} is our system, which dynamically creates and assigns resources to tasks.
This is the same as \variant{\sys{}-static}, except that we replace the round-robin approach for assigning tasks with the scheduling policy described in \Cref{sec:scheduler}.
This achieves 91.0\% of the optimal run time based on maximum GPU throughput, and it achieves 1.95$\times$ better run time than \variant{\sys{}-microbatch}, as well as 2.6$\times$ and 1.29$\times$ better than Flink and \variant{\sys{}-static}, respectively.

\paragraph{Takeaways:} (1) Dynamic task creation promotes memory-efficient pipelining, (2) \sys{}'s dynamic resource assignment further improves throughput over static assignment.

\subsubsection{Fault tolerance in heterogeneous clusters}
\label{sec:eval:ft}

We use the same workload as in \Cref{sec:eval:video} to demonstrate \sys{}'s ability to scale using clusters with heterogeneous node types and to smoothly scale clusters up and down, including during failures.
\Cref{fig:video:tput} shows the throughput during failure recovery when processing 10\% of the dataset using a cluster with 1 \inst{g5.xlarge} node (4 vCPU, 1 GPU) and 1 \inst{m7i.2xlarge} node (8 vCPU).
For all systems, we inject a node or executor failure at $t$=15min.
Executor failure kills one worker process.
Node failure disconnects the CPU-only node and reconnects it at $t$=30min.
We compare \sys{} (solid curves) against \sys{} modified to emulate the global checkpoint strategy used in many systems with static task assignment. The modified \sys{} (dashed curves) takes an empty checkpoint every 6min.

With checkpointing, both executor and node failures force a global rollback, causing downtime until the system recovers the lost work.
Downtime under node failure is longer because the system has fewer resources while the CPU-only node is disconnected.
Adding the CPU-only node back at $t$=30min forces another global rollback.
Meanwhile, \sys{} has no noticeable throughput drop under executor failure, and scales smoothly as nodes are removed and added.

\paragraph{Takeaways:}
Compared to systems that use static task placement and global checkpointing, \sys{} achieves similar run-time overhead. Meanwhile, \sys{} also smoothly handles CPU failures and cluster reconfiguration.

\subsection{Training: \sys{} vs.~ML data loaders}
\label{sec:eval:training}

\subsubsection{ResNet Training}

We run the ResNet-50 ImageNet training benchmark from MLPerf~\cite{mlperf-training}.
The data preprocessing pipeline loads images from local disk (\variant{local}) or cloud storage (\variant{S3}), decodes, and randomly crops and flips the images.
We compare training throughput vs.~tf.data on a \inst{g5.2xlarge} VM.
We do not measure PyTorch DL, as tf.data showed comparable or better results for the same benchmark~\cite{tfdata}.

\Cref{fig:eval:image-training} shows training throughput over time.
tf.data executes data preprocessing using a pool of worker threads running in each GPU trainer process.
Thus, the job fate-shares with \emph{any} preprocessing task that fails due to out-of-memory (OOM).
When reading data from local disk, tf.data's throughput is 19\% lower than \sys{}'s because a lower batch size was required to prevent OOM failures.
Meanwhile, \sys{} is able to complete because GPU trainer failures are isolated from CPU worker failures, and CPU workers can be respawned in seconds, without impacting pipeline throughput~(\secref{sec:eval:ft}).

When reading data from S3, tf.data is 88\% slower than the max GPU throughput because S3 loading is the bottleneck and the GPU's local CPUs are insufficient.
Meanwhile, \sys{} (S3) adds a CPU-only \inst{m7i.2xlarge} node to scale out S3 loading independent of the GPU trainers.
This achieves 93\% of the maximum GPU training throughput.

\paragraph{Takeaways:} Compared to single-node ML data loaders, \sys{} offers: (1) failure isolation between heterogeneous resources, and (2) ability to leverage heterogeneous node types.

\newsavebox{\imagetrainingbox}
\begin{figure*}[t]
    \sbox{\imagetrainingbox}{\includegraphics[width=0.25\textwidth]{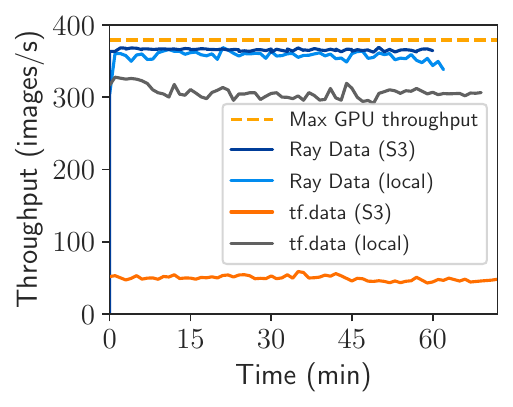}}%
    \begin{subfigure}[b]{\wd\imagetrainingbox}
         \raisebox{-6pt}[\dimexpr\ht\imagetrainingbox-6pt\relax][0pt]{\usebox{\imagetrainingbox}}
        \caption{}
        \label{fig:eval:image-training}
    \end{subfigure}\hfill
    \begin{subfigure}[b]{0.73\textwidth}
      \parbox[b][\ht\imagetrainingbox][c]{\linewidth}{\small
        \setlength{\tabcolsep}{0pt}
        \begin{tabular*}{\linewidth}{@{\extracolsep{\fill}}lcccc}
        \toprule
        & Resources & Images/s & Run time (hours) & Total cost \\\midrule
        PyTorch DL (stream) & $4\times$ \inst{p4de.24xlarge} & 2,811 & 111.3 & \$18,192 \\\midrule
        \variant{\sys{}-staged} (batch) & $4\times$ \inst{p4de.24xlarge} & 0, then 4,068 & 90.3 \textbf{(-19\%)} & \$14,753 \textbf{(-19\%)} \\\midrule
        \sys{} (dynamic task & $4\times$ \inst{p4de.24xlarge} & 4,075 & 76.8 \textbf{(-31\%)} & \$16,275 \textbf{(-11\%)} \\
        creation and assignment) & $40\times$ \inst{g5.2xlarge} & & & \\
        \bottomrule
        \end{tabular*}}
      \caption{}
      \label{eval:training:stablediffusion:results}
    \end{subfigure}
    \caption{
      \textbf{(a)} Training ResNet-50, loading data from local disk vs.~cloud storage (S3). All experiments run on 1 \inst{g5.2xlarge} node (1 NVIDIA A10G GPU, 8 vCPUs), except for \sys{} (S3), which uses an additional \inst{m7i.2xlarge} (8 vCPUs) node to scale out S3 loading.
      \textbf{(b)} Run time and cost for one epoch of Stable Diffusion pre-training.
      }
\end{figure*}

\subsubsection{Pre-Training Stable Diffusion}
\label{sec:eval:stablediffusion}

Pre-training of large multimodal models is one of the most demanding heterogeneous workloads.
We run the Stable Diffusion (SD) pre-training pipeline shown in \Cref{fig:stablediffusion} and compare different execution modes.
We execute 1 training epoch over a dataset of 2B images, on a cluster of $4\times$ \inst{p4de.24xlarge} nodes, with 8 A100 GPUs each.
This pipeline requires both CPUs and GPUs for data preprocessing:
(1) \code{loadText/loadImage+clip} (CPU): Load and preprocess pairs of image and text, (2) \code{Encoder} (GPU): Use pre-trained encoder models for image and text to produce embeddings, and (3) \code{UNet.train()} (GPU): Train SD, using PyTorch fully-sharded data parallelism (FSDP)~\cite{fsdp}.

\begin{figure}[t]
      \centering
      \includegraphics[width=0.38\textwidth,trim=0.2cm 0cm 0.1cm 0cm, clip]{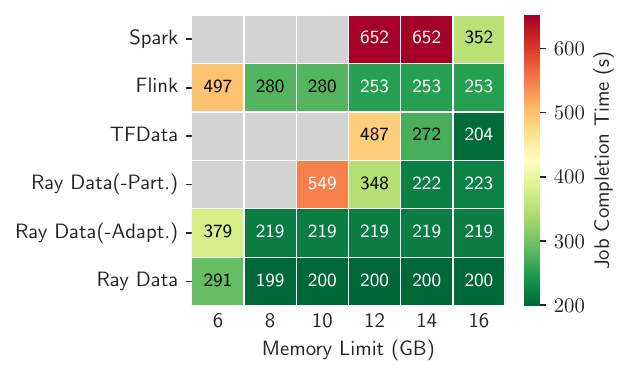}
      \caption{
      Run times for systems under different memory limits on a synthetic benchmark. Grey means the system is unable to finish due to OOM. \variant{\sys{}(-Part.)} means \sys{} without dynamic repartition~(\secref{sec:system:repartition}). \variant{\sys{}(-Adapt.)} means \sys{} without adaptive memory-aware scheduling~(\secref{sec:scheduling:impl}).
      }
      \label{fig:eval:synthetic-heatmap}
\end{figure}

\Cref{eval:training:stablediffusion:results} shows training throughput and total cost.
PyTorch DL is a data loader custom-built for PyTorch that statically assigns tasks to a process pool.
Its throughput is lowest because \code{Encoder} preprocessing competes with trainers for GPU memory.
\variant{\sys{}-staged} runs data preprocessing as an offline job and stores precomputed embeddings in cloud storage, which are then loaded back into memory during training.
This is preferable if the same embeddings are used multiple times. \variant{\sys{}-staged} achieves 19\% higher throughput than PyTorch DL because \code{UNet} can use more GPU memory and therefore a larger batch size.

\sys{} runs all data preprocessing concurrently with training.
This is preferable if using random transforms or for iterative development.
\sys{} also disaggregates
\code{Encoder} and \code{UNet}, placing \code{Encoder} onto cheaper A10G GPUs (\inst{g5.2xlarge}).
This results in 31\% better throughput than PyTorch DL, because \code{UNet} has full GPU resources, and 15\% better throughput than \variant{\sys{}-staged}, because embeddings are kept in memory.

\paragraph{Takeaways:} Compared to existing ML data loaders, \sys{} can: (1) support both staged and pipelined preprocessing, and (2) disaggregate operators using remote CPUs and heterogeneous GPUs to boost overall throughput.

\subsection{Microbenchmarks}
\label{sec:eval:mb}

\subsubsection{Memory-aware pipelining}
\label{sec:eval:mb:memory}

We evaluate how the various systems schedule heterogeneous pipelines with and without memory pressure.
The 3-stage pipeline is:
    (1) \op{Load} (CPU): 160 tasks, each producing 5 100\,MB rows after 5s,
    (2) \op{Transform} (CPU): sleep for 0.5s per row, then return a new 100\,MB row,
    (3) \op{Inference} (GPU): 0.5s per row.
We use 1 \inst{m6i.2xlarge} node with 8 vCPUs, 4 simulated GPU slots, and 32\,GB RAM.
The theoretical best job completion time with unlimited memory is $(160\times 5s + 800\times 0.5s) / 8= 150s$.
We compare against systems using static vs.~dynamic task creation and placement.

\Cref{fig:eval:synthetic-heatmap} shows job completion time vs.~total memory limit.
We limit memory through system-specific configurations, e.g., executor memory for Spark.
We use POSIX \code{rlimit} to verify that each system respects its memory limit, and tune each system's parallelism (e.g., executor count) if not.

Spark materializes all data between stages, achieving at best $2.35\times$ optimal run time.
At 12--14GB memory, Spark must use fewer executors resulting in $4.34\times$ the optimal run time, and at lower memory limits, Spark OOMs.
This is because Spark uses static partitioning, which is sensitive to memory pressure~(\secref{sec:bg:batch}).

Flink is less sensitive to the memory limit than Spark and achieves up to $1.68\times$ optimal.
This is because executors dynamically batch outputs and apply backpressure to avoid OOM~(\secref{sec:bg:stream}).
At lower memory limits, Flink must run fewer executors because each CPU slot is multiplexed among multiple operator threads, resulting in up to $2\times$ worse throughput.

tf.data offers an adaptive scheduler and memory budget, similar to \sys{}, albeit non-distributed.
However, we found that the memory budget was not always enforced, requiring manual tuning of the thread count.
tf.data achieves the same throughput as \sys{} at 16\,GB memory limit, but OOMs at lower memory limits.

\sys{} achieves $1.3\times$ the optimal run time at all memory limits except the lowest.
We also conduct ablation studies. \variant{\sys{}(-Part.)} disables \sys{}'s dynamic repartition~(\secref{sec:system:repartition}), resulting in too-large partitions similar to Spark.
\variant{\sys{}(-Adapt.)} uses \sys{}'s pessimistic policy instead of its optimistic policy~(\secref{sec:scheduling:impl}), resulting in 10--30\% worse performance than \sys{}.
\sys{} is also less sensitive than Flink to memory pressure because the system explicitly time-slices executors, instead of using multithreading.

\paragraph{Takeaways:} For heterogeneous applications under memory pressure, batch processing systems are unstable. \sys{} is as stable as stream processing systems, due to its dynamic repartition, and it can further leverage run-time profiling.

\subsubsection{Overhead of partitioning}
\label{sec:eval:mb:partitioning}

\sys{} manages distributed partitions with a centralized scheduler.
\Cref{fig:eval:partition-throughput} evaluates \sys{}'s system overhead by measuring throughput vs.~partition size on a 2-stage synthetic pipeline.
We use $8192\times$1\,MB input rows, and simulate 10\,ms processing time per row per stage.
The smallest partition sizes incur overhead from RPCs and bookkeeping, and the largest result in poor load-balancing.
To strike a balance, \sys{}'s default target partition size is 128\,MB.

\subsubsection{Scalability}
\label{sec:eval:mb:scalability}

To evaluate the scalability of \sys{}, we run an empty workload which creates then consumes a large dataset on the cluster.
As there is no computation to amortize per-task overhead, the scheduler must handle sustained high task churn.
We use 1 \inst{m8i.4xlarge} (16 vCPU, 64GiB RAM) as the head node and up to 32 \inst{m8i.2xlarge} (8 vCPU, 32GiB RAM) worker nodes.
The size of the dataset is proportional to the number of nodes in the cluster, with 5GB per node.
We compare \sys{} against two baselines: \variant{Ray} and \variant{Ray Generator}.
\sys{} uses the default partition size of 128MB.
\variant{Ray} spawns tasks that each return 128MB of data, then immediately passes each result to a consumer task.
\variant{Ray Generator} does the same, except that it uses the generator tasks described in \Cref{sec:system:repartition}.

\begin{figure}[t]
  \centering
  \begin{subfigure}[b]{0.23\textwidth}
      \centering
      \includegraphics[width=\textwidth,trim=0.1cm 0cm 0cm 0cm, clip]{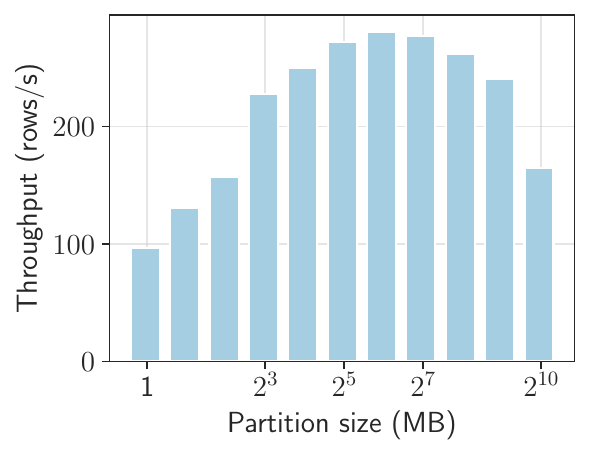}
      \caption{}
      \label{fig:eval:partition-throughput}
  \end{subfigure}
  \begin{subfigure}[b]{0.235\textwidth}
      \centering
       \includegraphics[width=113.10bp,trim=0.1cm -1.55bp 0cm 0cm, clip]{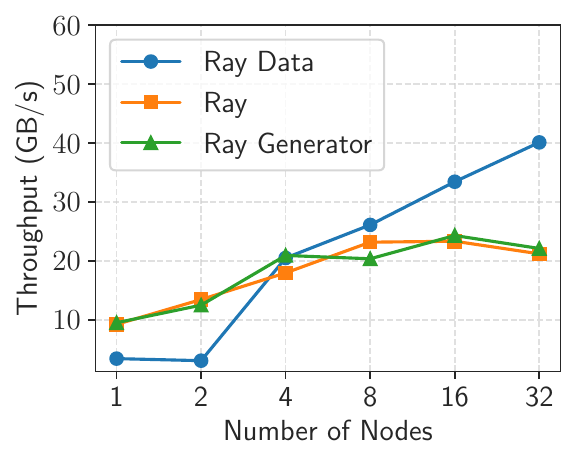}
      \caption{}
      \label{fig:eval:scalability}
  \end{subfigure}
  \caption{\textbf{(a)} Effect of partition sizes on throughput in \sys{}. \textbf{(b)} Scalability experiment showing throughput over number of nodes compared to using raw Ray tasks or generator tasks (\variant{Ray Generator}).}
\end{figure}

\Cref{fig:eval:scalability} shows the throughput versus the number of nodes in the cluster.
\variant{Ray Generator} achieves similar throughput to \variant{Ray}; this validates that our extensions to Ray to support generator tasks~(\secref{sec:system:repartition}) add negligible overhead.
\sys{} achieves worse throughput than \variant{Ray} and \variant{Ray Generator} on clusters with $\le2$ nodes because of warmup time from its additional query planning step~(\secref{sec:system:planning}).
However, \sys{} scales linearly and delivers up to 1.8$\times$ better throughput than \variant{Ray} and \variant{Ray Generator} on larger clusters.
This is due to its application-aware scheduling policy~(\secref{sec:scheduler}), which uses the logical dataflow graph~(\Cref{fig:applications}) to improve load balancing and avoid unnecessary spilling.

\section{Additional Related Work}

\paragraph{Data loaders for ML training.}
Both PyTorch~\cite{paszke2019pytorch} and TensorFlow (tf.data~\cite{tfdata}) provide map-style data loaders for ML training.
tf.data
automatically shards the dataset, while PyTorch DL requires the user to shard the dataset themselves.
Both are single-node systems colocated with a GPU trainer and share similar limitations: they cannot leverage CPU-only nodes, load-balancing must be done by sharding before execution, and the training job fate-shares with the data loader.

tf.data service~\cite{audibert2023tf} improves upon these by enabling disaggregated CPU-based preprocessing for training.
\sys{} further provides exactly-once semantics (as opposed to at-most-once), batch inference support, and disaggregation with arbitrary heterogeneous nodes in addition to CPU-only nodes.
Several recent works propose query optimizers for training preprocessing~\cite{graur2022cachew,zhao2024cedar}. These are complementary to \sys{}.

\paragraph{Scheduling for resource heterogeneity.}
The scheduling problem described in \cref{sec:scheduler} is most similar to the generalized processor sharing~\cite{gps} problem.
Our solution is inspired by the weighted fair queuing algorithm~\cite{bennett-zhang,demers}.
The differences are (1) the flows in network scheduling are independent of each other, whereas operators in a data pipeline have dependencies, (2) we consider multiple resource types, and (3) the packet processing time is usually fixed, whereas data operator processing times are unpredictable.

Recent scheduling works attempt to adapt fair queuing and autotuning to heterogeneous resource environments.
Dominant resource fair queuing~\cite{drfq} addresses the problem of (2) but not (1) or (3).
tf.data~\cite{tfdata} introduces an autotuning algorithm that uses gradient descent to find the parallelism for each operator that reduces end-to-end latency.
Such policies can also be integrated into \sys{}'s scheduler.

\section{Discussion}

In this work, we present an online scheduler,
but still make certain planning decisions statically, including the number of input partitions (\secref{sec:system:planning}).
The user also needs to specify the cluster shape.
While we designed \sys{} for heterogeneous workloads, we also observe that it enables a key opportunity in future data processing systems: dynamic query planning.
We envision a fully autotuning and autoscaling system that can jointly configure the application and cluster.

\paragraph{Conclusion.}
With the rise of LLMs, even modalities such as text may now require expensive deduplication~\cite{openaiembeddings}, retrieval~(\secref{sec:eval:rag}), GPU-based encoding, and joins with image or video data~(\Cref{fig:stablediffusion}).
Also, as the parallelism strategies used in inference and training pipelines become more complex, future data processing systems must support more flexible APIs for sharding and reducing data copies.

We believe that ML systems will continue to grow in the complexity of their data processing needs, as evidenced by trends such as test-time training~\cite{testtimetraining,gandelsman2022testtime}, RAG~\cite{rag}, and multimodal models~\cite{gemini,openai2024gpt4}.
To keep up with this demand, we must build more flexible, heterogeneity-aware, and scalable data processing systems.

\bibliographystyle{plain}
\bibliography{main}

\end{document}